\documentclass{article}
\usepackage{authblk}
\usepackage{upgreek}
\usepackage{arxiv}
\usepackage{amsmath}
\usepackage[utf8]{inputenc} % allow utf-8 input
\usepackage[T1]{fontenc}    % use 8-bit T1 fonts
\usepackage{hyperref}       % hyperlinks
\usepackage{url}            % simple URL typesetting
\usepackage{booktabs}       % professional-quality tables
\usepackage{amsfonts}       % blackboard math symbols
\usepackage{nicefrac}       % compact symbols for 1/2, etc.
\usepackage{microtype}      % microtypography
\usepackage{lipsum}
\usepackage{graphicx}
\graphicspath{ {./images/} }

\title{Polarized hyperspectral imaging with single fiber bundle via incoherent light transmission matrix approach}

\author[1]{Yitong Li}
\author[1]{Zhengbo Zhu}
\author[1]{Ze Li}
\author[1,2,*]{Donglin Ma}

\affil[1]{School of Optical and Electronic Information and Wuhan National Laboratory for Optoelectronics
Huazhong University of Science and Technology, Wuhan, Hubei 430074, China}
\affil[2]{Shenzhen Huazhong University of Science and Technology, Shenzhen 518057, China}

\affil[*]{Corresponding author: madonglin@hust.edu.cn}

\begin{document}
\maketitle
\begin{abstract}
The scattering of multispectral incoherent light is a common and unfavorable signal scrambling in natural scenes. However, the blurred light spot due to scattering still holds lots of information remaining to be explored. Former methods failed to recover the polarized hyperspectral information from scattered incoherent light or relied on additional dispersion elements. Here we put forward the transmission matrix (TM)
approach for extended objects under incoherent illumination by speculating the unknown TM through experimentally calibrated or digitally emulated ways. Employing a fiber bundle as a powerful imaging and dispersion element, we recover the spatial information in 252 polarized-spectral channels from a single speckle, thus achieving single-shot, high-resolution, broadband hyperspectral imaging for two polarization states with the cheap, compact, fiber-bundle-only system.
 Based on the scattering principle itself, our method not only greatly improves the robustness of the TM approach to retrieve the input spectral information, but also reveals the feasibility to explore the polarized spatio-spectral information from blurry speckles only with the help of simple optical setups.
\end{abstract}

% keywords can be removed

\section{Introduction}
\label{sec:Intro}
The light scattering is quite common in natural scenes. When seeing through rain, fog, cloud, or water, a hidden object behind the disordered medium is turned into a heavily blurred light spot, making the original object hard to be discriminated. However, the full information, especially polarized-hyperspectral information, is still encoded in the light spot. As a result, to recover the hidden information of an object under incoherent illumination from its speckle is highly rewarding.

In recent years, imaging \cite{SternKatz-8,Bertolotti2012,Katz2014,Zhuang2016,Edrei2016,Choi2012}, polarization detection \cite{Soni:16}, and spectrum analysis \cite{Redding2012,Redding2014,Kueruem2019,Redding2013,Bruce2019} techniques through complex media have been developed to extract object information behind scattering media and fibers. In the aspect of imaging, methods such as speckle correlation \cite{SternKatz-8,Bertolotti2012,Katz2014,Wu2016}, point-spread-function (PSF) deconvolution \cite{Tang2018,Zhuang2016,Edrei2016}, transmission matrix (TM) \cite{Popoff2010,Choi2011,Carpenter2013,Gu2015,Gordon2019,Choi2012}, and phase conjunction \cite{Yang2019,Papadopoulos2012,Si2012} have proved that the complex medium could be a powerful imaging device with large numerical aperture (NA) and high resolution \cite{Edrei2016}. In the aspect of spectrum analysis, by measuring the spectral intensity TM in a fixed spatial mode and polarization state, complex media such as fibers \cite{Redding2012,Redding2013,Liew2016} and scattering media \cite{French2017} show excellent spectral resolution to build a broadband compact spectrometer. Deep learning techniques are also deployed to recover the object from interferometric or intensity-only speckles \cite{Ploeschner2015,rahmani2018multimode}.

However, those methods face trouble to recover the polarized hyperspectral information. Most of those methods only work with coherent, single-spatial-mode, or narrow bandwidth light. Although several methods have been found to perform hyperspectral imaging by dividing the field of view (FOV) in pixel-level \cite{French2018,Kueruem2019,French2017}, those techniques are still limited to objects under coherent illumination with a strict incident angle range. While deconvolution techniques perform well in color imaging \cite{Zhuang2016,Sahoo2017}, they always assume that PSFs associated with different spectral channels are uncorrelated with each other \cite{Li2019}. This assumption brings difficulties to recover the intensity of the whole continuous spectrum.

To recover the hyperspectral information for objects under incoherent illumination, some techniques have been proposed recently with the use of the coded aperture \cite{Li2019}, multiple scattering materials \cite{Muskens2018,French_2018}, or the spectral filter array \cite{Monakhova_2020}. Those methods encode the speckle of the scattered object and recover the hyperspectral information through PSFs' library \cite{Anand2020}. However, those techniques still rely on conventional dispersion and elements such as spectral filters or prisms. So the spectral resolution is restricted, and the scattering medium's dispersion ability is not taken advantage of, even become a limitation itself. Besides, those methods also fail to recover the polarization information from speckles.

Compared with conventional imaging and dispersion elements, fibers have superior properties in cheapness, compactness, and high spectral resolution. Fibers are a typical kind of disordered or scattering media, whose spectral intensity TM,  however, can be evaluated only in the single spatial mode by far \cite{Redding2014}, resulting in a fundamental contradiction with natural hyperspectral scenes. Moreover, it is also hard to image an object under incoherent illumination with a single multi-mode fiber. Compared with a single fiber, a fiber bundle has the 'memory effect' feature \cite{Porat2016} just as scattering media \cite{Katz2012} . So the fiber bundle can become the potential choice to exceed the research gap.

In this paper, we propose and demonstrate our single-shot polarized hyperspectral imaging method with an all-fiber-designed optical system and a monochrome camera. The speckle pattern is more sensitive to wavelength or polarization state than the object's spatial distribution. With this advantage, we deduce the convolution approach of the spectral intensity TM in incoherent situation and put forward methods to approximate the unknown TM. With our methods, the spatial, spectral, and polarization information can be separately recovered. Based on the approach without any assistance of additional imaging, polarization or dispersion elements, we perform hyperspectral imaging through the disordered medium from 400nm to 650nm in 2 orthogonal polarization states, adding up to 252 imaging channels with 2nm spectral resolution. Since the fiber bundle promises to have large spectral bandwidth and NA, those system's properties can be further improved by changing the calibration range. We also study factors that affect the hyperspectral imaging quality and propose the method to solve the mismatch problem between the object and the calibration speculation.

\section{WORKING PRINCIPLE}
\label{secII}

At a conceptual level, the polarized hyperspectral imaging system consists of a target object, dispersion elements, imaging elements, polarization elements and a detector. Here we use a fiber bundle as the only dispersion, polarization and imaging element. The fiber bundle consists of multimode core fibers arranged in a fixed spatial arrangement, and its conventional use is to transport an image from one end to another. However, when the object plane (OP) and the image plane (IP) are away from the bundle's facet, the output image will be turned into a speckle. For an extended object, each spatially and/or temporally incoherent component creates its speckle pattern at the IP, so the detector receives the overlaid, blurry-looking light spot with a low signal-to-noise ratio (SNR). This is different from the object under coherent illumination where its speckles have high contrast and sharp grids.

\begin{figure}[htbp]
\centering
{\includegraphics[width=0.7\linewidth]{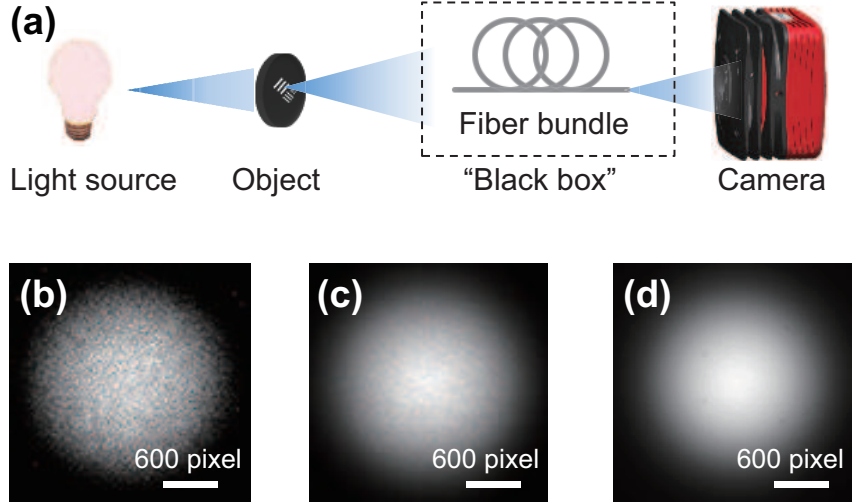}}
\caption{(a) Hyperspectral imaging system based on a fiber bundle. An incoherent light source illuminates the object. (b) The system's speckle when the narrow linewidth light illuminates the point object: a 2nm linewidth light source illuminates a 0.1mm pinhole. Here a pinhole that is smaller than the system's diffraction limit is treated as a point object. (c) The speckle when the narrow linewidth light illuminates an extended object: a 2nm linewidth light source illuminates a 2mm aperture. (d) The speckle when the broadband light illuminates the extended object: a 10nm linewidth light source illuminates the 2mm aperture. The speckle's contrast decreases greatly as the incident light's incoherent components become more complex.}

\label{fig1}
\end{figure}
\begin{figure*}[ht]
\centering
%{\includegraphics[width=\linewidth]{Fig/Fig2-complex.eps}}
{\includegraphics{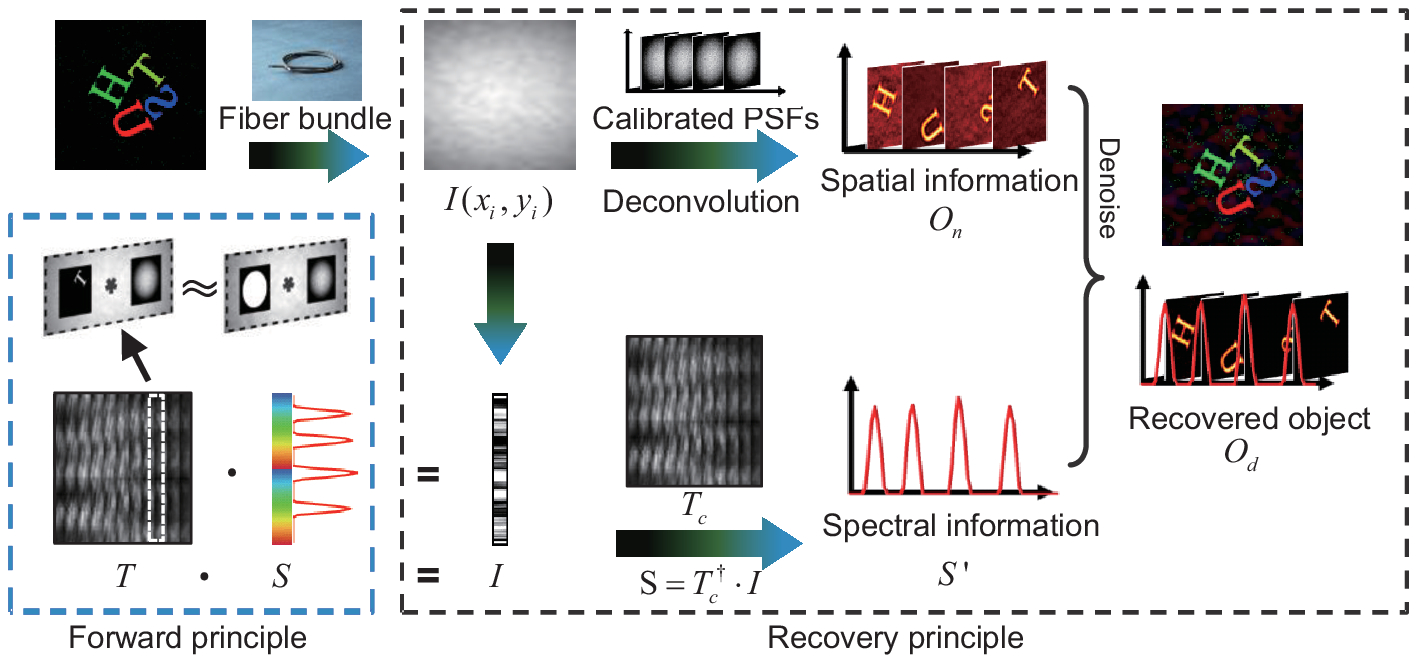}}
\caption{ The data processing pipeline. In the memory effect range, the object's speckle $I(x_i,y_i)$ can be extracted as the convolution of PSFs and object $O$. So we apply the TM to describe the information scrambling and replace the unknown TM with the calibrated result, then inverse the TM to calculate the spectrum. We also perform the deconvolution method in each channel to obtain the spatial intensity distribution and perform denoising with the recovered spectrum to reconstruct the real object. }
\label{fig2}
\end{figure*}

Figure  \ref{fig1} shows the concept of our hyperspectral imaging system. Generally speaking, analyzing the propagation of incoherent light through fiber bundle by mode theory can be a tough task, but we can simplify the situation to a black box system. Figure \ref{fig2} shows our method. The basic idea is that each point of the object generates an speckle pattern at the IP, which is the point spread function (PSF). So, the forward hyperspectral imaging principle in single polarization can be described as:
\begin{equation}
I({{x}_{i}},{{y}_{i}})=\int{PSF({{x}_{i}},{{y}_{i}};{{x}_{o}},{{y}_{o}},{{\lambda }_{o}})\cdot O({{x}_{o}},{{y}_{o}},{{\lambda }_{o}})d{{x}_{o}}d{{y}_{o}}d{{\lambda }_{o}}}
\label{eq:eq1}
\end{equation}
where the subscripts $i$ and $o$ denote the IP and the OP respectively, and ${\lambda }_{o}$ denotes the wavelength making up the object. Within the memory effect range, we have $PSF({{x}_{i}},{{y}_{i}};{{x}_{0}},{{y}_{0}},{{\lambda }_{0}})=PSF({{x}_{i}},{{y}_{i}})$, and the forward imaging procedure becomes convolution imaging:
\begin{equation}
I({{x}_{i}},{{y}_{i}})=\int{PSF({{\lambda }_{o}})*O}({{\lambda }_{o}})d{{\lambda }_{o}}
\label{eq:eq2}
\end{equation}

we can rewrite Eq. \ref{eq:eq2} from integral form to matrix form by spectral discretization to separate the spatial and spectral information of the object:
\begin{equation}
I({{x}_{i}},{{y}_{i}})=T\cdot S({{\lambda }_{o}})
\label{eq:eq3}
\end{equation}
where $S$ represents the column vector of the object's spectrum, each column of $T$ represents the speckle image $PSF({{\lambda }_{o}})*O({{\lambda }_{o}})$ in each spectral channel. In Eq. \ref{eq:eq3}, each spectral channel is centered at ${\lambda}_{o}$  and spaced by $\delta\lambda$ with the spectral width ${D}\lambda$. We assume that the spectral correlation range of the system is  $\Delta\lambda$ , and $\delta\lambda$ should follow $\delta\lambda\le\max(\Delta\lambda,D\lambda)$ to make sure that the spectral channels are consequent. While $O$ and $PSF$ are unknown and we cannot directly calculate or measure the incoherent TM, we notice that the blurry speckle is actually the overlaid pattern of PSF, thus giving us the idea to approximate the TM through a calibration. Here we use a round aperture $O_c$ to substitute the object. To ensure the $PSF*O_c$ and $PSF*O$ similar enough, we assume that $O_c$ and $O$ are at the same location at the OP, and the diameter of $Oc$ should follow:
\begin{equation}
{{D}_o}\le {{D}_c}\le {{D}_{ME}}
\label{eq:eq5}
\end{equation}
where ${D}_o$ is the diameter of object, ${D}_c$ is the diameter of $O_c$, and ${{D}}_{{ME}}$ is the memory effect range.

During the calibration, we utilize a wavelength-shifted incoherent light source to illuminate the aperture. Firstly, we assume the full width at half maximum (FWHM) of the light source's spectrum to be ${D}{\lambda}\ll\delta\lambda$, so we can treat the light source as single wavelength. Then the speckle intensity $I_{\lambda_k}$  at the scanning wavelength $\lambda_k$ follows:

\begin{equation}
I_{\lambda_k}={PSF({{\lambda }_{k}})*O_c}
\label{eq:eq4}
\end{equation}
where $k$ refers to the $k_{th}$ step of wavelength scanning. Then we reshape $I_{\lambda_k}$ into a column vector as the $k_{th}$ column of ${{T}_{{c}}}$.  On condition that ${D}\lambda\sim\delta\lambda$, the linewidth of the calibration source should be considered. At the $k_{th}$ scanning, the scanning light has the spectrum ${S_k}$ with the central wavelength ${\lambda}_k$. So the reshaped speckle forms the $k_{th}$ column of the broad-linewidth calibration matrix ${{T}_b}$, and $S_k$ forms the $k_{th}$ column of the tansformation matrix $S_c$. The relationship between ${{T}_c}$ and ${{T}_b}$ follows:
\begin{equation}
{{T}_{b}}={{T}_{c}}\cdot {{S}_{c}}
\label{eq:eq7}
\end{equation}
So $T_c$ is calculated by:
\begin{equation}
{{T}_{c}}={{T}_{b}}\cdot {{S}_{c}}^{\dagger}
\label{eq:eq8}
\end{equation}
where ${{S}_{c}}^{\dagger}$ denotes the pseudo-inversing of ${S}_{c}$. After the calibration, we replace TM with the calibrated $T_c$:
\begin{equation}
I({{x}_{i}},{{y}_{i}})\approx{{T}_{c}\cdot S({{\lambda }_{o}}})
\label{eq:eq6}
\end{equation}
Then we can inverse $T_c$ to recover the input spectrum:
\begin{equation}
S'\approx{{T_c}^{\dagger}\cdot {I}}
\label{eq:eq9}
\end{equation}
where $S'$ is the recovered spectrum. The matrix inversion would face with ill-conditions. Here we adopt a matrix pseudo-inversion algorithm based on singular value decomposition. Further optimization is an option, such as employing nonlinear optimization procedure to minimize $\left\| I-T_c\cdot S' \right\|_{2}^{2}+{{\gamma }_{1}}{{\left\| S' \right\|}_{1}}$, where $S'$ is the recovered spectrum taken as the initial value. However, this optimization procedure costs much time due to the large dimensions of ${{T}_{c}}$ and has a limited effect.

After the input spectrum is solved, we deconvolute the speckle with each $PSF$ to acquire the spatial intensity distribution in each channel:
\begin{equation}
{{O_d}}\left( {{\lambda }_{o}} \right)=deconv\left( I,PSF\left( {{\lambda }_{o}} \right) \right)
\label{eq:eq10}
\end{equation}
Here we conduct a standard deconvolution operation with the Wiener filter algorithm in MATLAB. However, due to the correlation of PSFs between different $\lambda$ and noise in the image, the deconvolution results only represent the relative spatial intensity distribution in each channel and are affected by strong background noise. Generally, there are channels with nearly zero spectral intensity to estimate background noise. For the background spectral channel at ${{\lambda }_{1,\ldots,k,\ldots,n}}$, we have:
\[S'(\lambda_{k})\approx0\]
And
\begin{equation}
BG=\frac{1}{n}\underset{i=1}{\overset{n}{\mathop \sum }}\,{{O}_{n}}\left( {{\lambda }_{k}} \right)
\label{eq:eq11}
\end{equation}
Then we denoise $O_n$ with $S'$ and $BG$:
\begin{equation}
{{O}_{d}}({{\lambda }_{i}})=S'({{\lambda }_{i}})\cdot \frac{{{O}_{n}}({{\lambda }_{i}})-BG}{\sum\limits_{{{x}_{o}},{{y}_{o}}}{({{O}_{n}}({{\lambda }_{i}})-BG)}}
\label{eq:eq12}
\end{equation}

On condition that the object mismatches the calibration aperture heavily, we put forward a method to rebuild the TM with a more exact estimation:
\begin{equation}
{{T}_{e}}( {{x}_{i}},{{y}_{i}};{{\lambda }_{o}})=PSF({{x}_{i}},{{y}_{i}};{{\lambda }_{o}})*{{O}_{d}}({{\lambda }_{o}})
\label{eq:eq13}
\end{equation}
Compared with the experimentally calibrated $T_c$, the emulated $T_e$ is much closer to the ground truth $T$, but the emulation procedure is heavily affected by noise in PSFs. Then $T_e$ is inversed just like $T_c$ in Eq. \ref{eq:eq9} to calculate the spectrum.

By now we have recovered the spectral and spatial information of the object for one polarization state. For two orthogonal polarization states, we should calibrate $T_c$ and $PSF$s in each polarization state. Then the whole TM can be cascaded by:
\begin{equation}
T_c = [T_{c1},T_{c2}]
\label{eq:eq14}
\end{equation}
where $T_{c1}$ and $T_{c2}$ are the calibrated TM in two orthogonal polarization states. Then we calculate $S'$ by Eq. \ref{eq:eq9} and cut off $S'$ by half to obtain the spectrum in each polarization state. After that, we perform the same imaging procedure in each polarization as we discussed above to retrieve the object’s intensity. As a result, the hyperspectral information for two polarization states is recovered.

\begin{figure}[htbp]
\centering
{\includegraphics[width=0.7\linewidth]{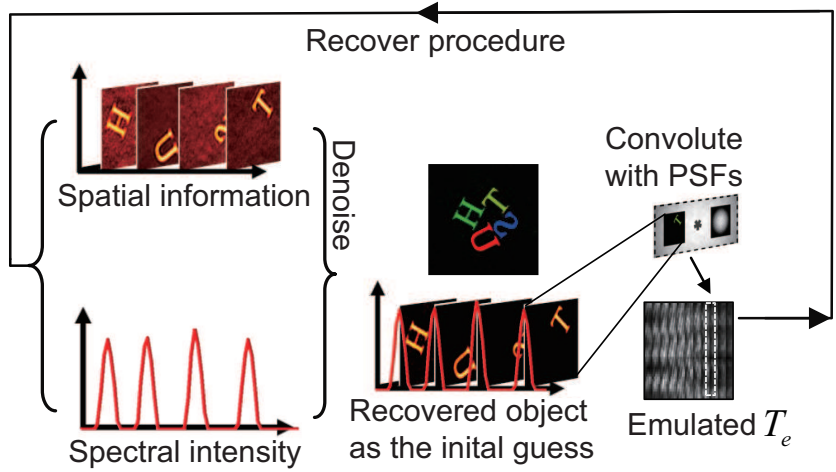}}
\caption{Approximating the TM with emulation speckles. Using $O_d$ as the initial guess, we emulate each channel's speckle pattern to achieve a more accurate approximation for the TM. The algorithm can be performed many times until the best result.}
\label{fig3}
\end{figure}

\section{METHODS/ IMPLEMENTATION DETAILS}
\label{sec:method}
\begin{figure}[tbp]
\centering
{\includegraphics[width=0.7\linewidth]{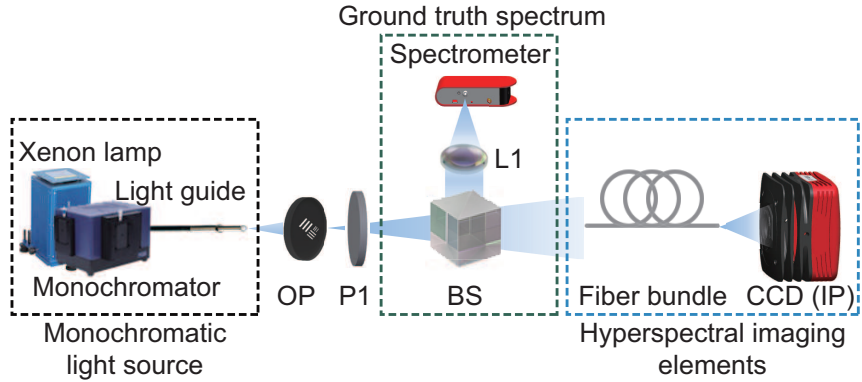}}
\caption{Experiment setup of the hyperspectral imaging system. The fiber bundle and the CCD are fixed during the whole experiment. A beam splitter BS, a lens L1, and a spectrometer are behind the polarizer to collect the spectrum as the true value. After the calibration, we change the light source as well as the object to perform hyperspectral imaging in different scenes.}
\label{fig4}
\end{figure}

Based on our demonstration, our hyperspectral imaging system only consists of a fiber bundle and a monochrome image sensor. The system setup is shown in Fig. \ref{fig4}. We build our system with an 8cm long multimode fiber bundle (Nanjing Chunhui Conventional imaging fiber bundle) with NA=0.6 and diameter ${{{D}}_{{bundle}}}$=2mm. It has 5500 cores with their core diameters ${{{d}}_{{core}}}$ varying from 12$\upmu$m to 15$\upmu$m. Its spectral decorrelation range is 2nm. To calibrate the system in our target spectrum range, we use a Xenon lamp (Perfect Light, 300W) combined with a monochromator (7IMSES) to form a tunable calibration light source. The FWHM of the source here is set to 2nm to calibrate each spectral channel one by one.

The light source is first coupled to a liquid fiber guide (Edmund \#1511438) to be homogenized spatially and then guided to the object. Here we use a 2mm aperture to calibrate $T_c$. The distance between the OP and the bundle's incident end is $U$=440mm. A monochrome CCD camera (Thorlabs 8051M-USB-TE) is placed $V$=30mm away from the bundle's exit end to collect the overall speckle. A 1:9 non-polarized beamsplitter together with a collecting lens and a spectrometer (OtO Photonics, 400nm-700nm) is behind the object to collect the actual value of the calibration spectrum as $S_c$. We locate the whole experiment setup in the black room to reduce background light and stray reflections. We also fix the fiber bundle in our tube lens system and couple it with the C-mount connector to our CCD to minimize the disturbance of the fiber bundle. After that, We correct the rotation angle of the fiber bundle to make sure that no image is rotated after transporting through the fiber bundle.

At the start of the calibration, a linear polarizer is placed behind the OP. In each orthogonal linear polarization state, speckles at different $\lambda_i$ are collected while the light source sweeps from 400nm to 650nm at an interval of 2nm. Simultaneously, the spectrum $S(\lambda_i)$ at each interval is also collected by the spectrometer. After calibrating $T_c$, we substitute the aperture with a 0.1mm pinhole. We perform a similar calibration procedure to acquire PSFs of the system at different wavelengths.

The target objects consist of resolution targets, figures, letters, images with their diameters varying from 1.3mm to 3mm. In the experiment, the target object is placed at the aperture's exact position to minimize the mismatch. Different light sources (LEDs, Xenon lamp, Lasers) are combined with filters and a polarizer to create an object with a different spectrum. Then we digitally add speckles together to create the speckle from a virtual object with complex spectral-spatial distribution.

We apply different data processing methods in the spectrum recovery and the imaging procedure. In the spectrum recovery, firstly each batch of 2x2 pixels in the raw speckle image is sampled into one super pixel to perform the spatial discretization with high fidelity while easing the compute stress. Then we crop the center 1200x1200 pixels of the speckle to remove most of background noise. The cropped images are processed by a median filter to remove thermal pixels. Then we smooth the speckle by a Gaussian filter with a standard deviation of 10 pixels. The filtered images are spatially normalized by dividing it by a low-pass-filtered version of itself (obtained by convolving the image with a uniform 5×5 kernel). The filtered speckle $I(x_i,y_i)$ of the calibration aperture at $\lambda_i$ is reshaped into a column vector to form the $i$-th column of $T_b$. We firstly inverse $S_c$ to calculate $T_c$ by Eq. \ref{eq:eq8}, then inverse $T_c$ to calculate $S'$ by Eq. \ref{eq:eq9}. Here we adopt a matrix pseudo inversion algorithm and treat the five smallest singular values of $T_c$ as zero.  Negative values in $S'$ are
assigned to zero. We also adopt the nonlinear regulation algorithm in CVX toolbox as optional \cite{cvx,gb08}. In the imaging procedure, the raw data of the speckle is processed by a median filter to conduct the Wiener deconvolution. The intensity of the recovered objects is then corrected by Eq. \ref{eq:eq12}.

\section{RESULT}
\subsection{Reconstruction of Every Single Spectral Channel}
\label{sec3a}
Here we first test our hyperspectral imaging system by reconstructing every spectral channel within the system bandwidth. We use a 1.3mm aperture as the target object and sweep the monochromator at an interval of 2nm. The linewidth is also set to 2nm. After the data collection is finished, each speckle is processed and reconstructed by Eq. \ref{eq:eq9} and the SVD algorithms.

\begin{figure}[!tb!]
\centering
{\includegraphics[width=0.7\linewidth]{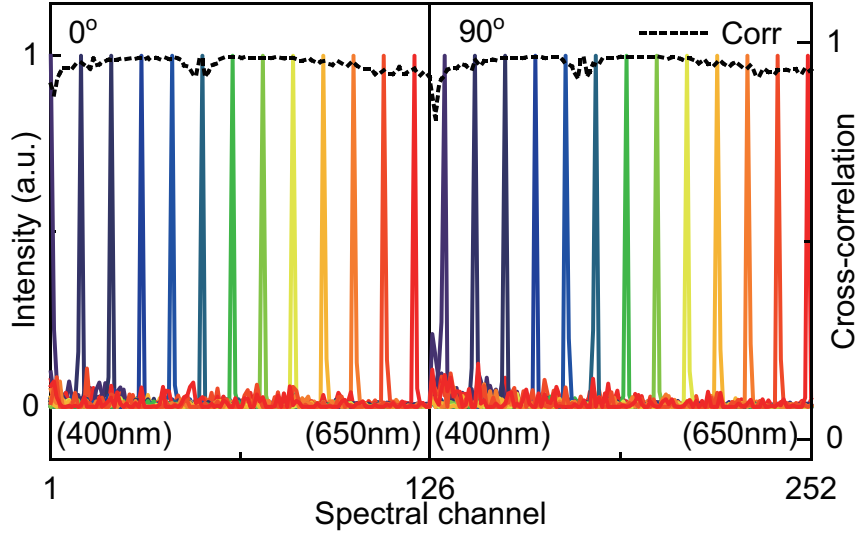}}
\caption{Reconstruction of spectral channels with 20nm spacing from 400nm to 650nm for two polarization states by the SVD algorithm. Each recovered channel is plotted in its spectral color, and the correlation index between the original spectrum and the reconstructed result is plotted as dashed line.}
\label{fig5}
\end{figure}

We plot 25 recovered channels in Fig. \ref{fig5} with its spectral colormap. The result shows that each spectral channel is reconstructed within about 2nm wavelength accuracy. The overlapping area at the lower intensity region shows that the reconstructed spectrum suffers from noise. The dashed line in Fig. \ref{fig5} shows the cross-correlation values between the ground truth and the recovered spectrum in 252 channels. The spectral channels near 400nm and 650nm show relatively poor reconstruction performance due to the low transmission efficiency of the coatings on the CCD and the fiber bundle near those wavelengths. Overall, the recovery quality is acceptable and can be further improved by adopting the nonlinear regulation algorithms.

We also perform the imaging procedure in each spectral channel. To evaluate the imaging quality, we calculate the structural similarity index measure (SSIM) relative to the object's binary image. It turns out to be 0.976 on average.

\subsection{Hyperspectral Imaging For Two Polarization States}

We start with a simple target object formed by a Jupiter figure and a rotated, mirrored letter '9'. Each sub-object has a 2nm spectral linewidth, a 2mm diameter, and an orthogonal polarization state.

\begin{figure}[!tb!]
\centering
{\includegraphics[width=0.7\linewidth]{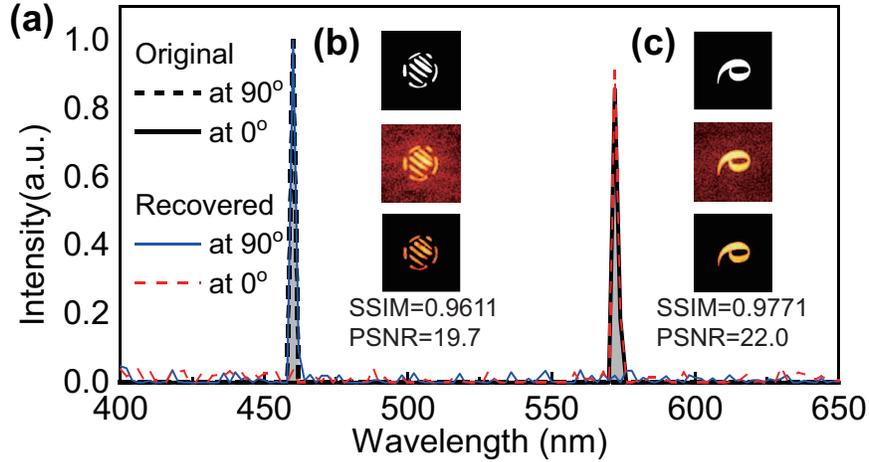}}
\caption{(a) The recovered and the original spectrum for two polarization states. (b),(c) The ground truth (top), deconvolution result $O_n$ (medium), and denoised image $O_d$ (bottom) in each spectral channel. Each image is normalized by its maximum intensity. The SSIM and PSNR values between the ground truth and the corrected images are also calculated. In our experiments, the rotated and the mirrored image results from the way to place the target object. Since the rotation effect has been corrected in Section \ref{sec:method}, no image is rotated or mirrored by the fiber bundle.}
\label{fig6}
\end{figure}

Figure \ref{fig6}(a) plots the original and recovered spectrums. The object's spectrum is recovered successfully with exact wavelength positioning. The relative spectral intensity is accurately recovered, and background noise is acceptable. Imaging results in Figs. \ref{fig6}(b) and \ref{fig6}(c) show that the denoising algorithm can significantly improve imaging quality. We also measure the recovered image's SSIM, peak signal to noise ratio (PSNR), and the recovered spectrum's correlation value corresponding to the ground truth. The correlation value between the original spectrum and the recovery is 0.9752. All those results show that our algorithms perform well in imaging and spectral recovery despite the difference between the calibration aperture and the hidden object.

\begin{figure*}[!t]
\centering
{\includegraphics[width=\linewidth]{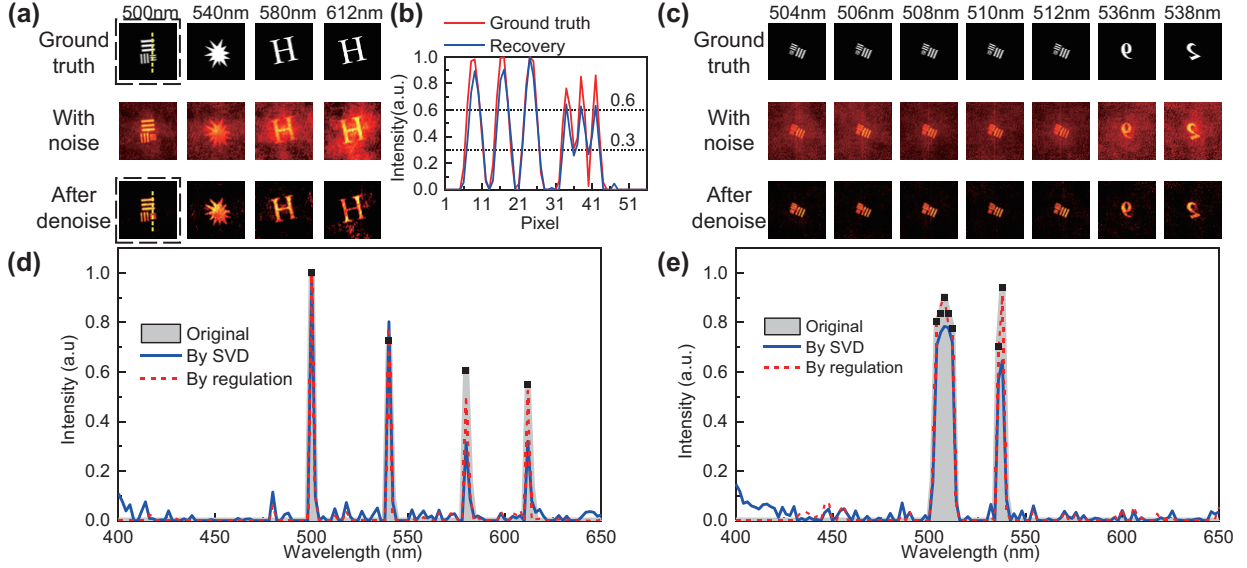}}
\caption{Hyperspectral imaging of objects with different dimensions and spectral sparsity.
(a) The ground truth and the recovered object at ${{90}^{\circ}}$ polarization state. Each Object in (a) has 2.5mm in diameter. (b) The profile of the resolution target in (a) at the white dashed line. (c) The ground truth and the recovered object at ${{0}^{\circ}}$ polarization state. Each Object in (c) has 1.3mm in diameter. (d),(e) The ground truth and the recovered spectrum at ${{90}^{\circ}}$ and ${{0}^{\circ}}$ polarization states. The black square at the original spectrum denotes the object's wavelength position and its intensity. Our regulation algorithm (dashed line) shows better results in the spectrum's contrast and intensity accuracy.  }
\label{fig7}
\end{figure*}

% While the object ‘9’ has low structure similarity with the calibration aperture, the system PSFs blurry the object into wavelength-related speckle pattern, making its speckle very similar to the calibrated speckle at its corresponding wavelength. In this blurry procedure, the blurry kernel, $PSF(\lambda)$ , contributes mostly to the speckle pattern feature, and the spatial distribution of the target object actually has less effect, as long as the sparsity of the object has similar magnitude with that of the calibration aperture.

\subsection{Hyperspectral Imaging of Objects with Different Dimensions and Sparsity.}

To further explain our hyperspectral imaging technique for various objects, we experiment with target objects with different dimensions and sparsity. We use a liquid crystal tunable filter and an LED to create the target object with 10nm spectral linewidth. Figure  \ref{fig7} plots our hyperspectral imaging result.

While our SVD algorithm recovers the spectrum with high wavelength's positioning accuracy, the reconstruction error occurs in the spectrum's intensity. Noise appears in the background channel and also causes intensity loss in the object's channel. The reconstruction error mainly results from low average SNR of the speckle's spectral channel.
The object with N spectral components has N times more complicated hyperspectral information. So the SNR of each spectral channel becomes 1/N of its original, which affects the matrix inversion procedure. As a result, the channel near 400nm or 650nm with initially lower calibration SNR is more seriously affected.
%some signal component of the original speckle turned into background noise in the recovered spectrum, causing errors in relative intensity recovery.
To improve the recovery quality, we conducted the nonlinear regulation optimization algorithm and set ${\gamma }_{1}$ = 0.3 in our experiment, proving satisfactory to correct the reconstruction errors.

We also test our system corresponding to the spatial-spectral resolution. While the wavelength's positioning accuracy is within 2nm, we set two different objects in 536nm and 538nm, and the hyperspectral imaging result in Fig. \ref{fig7}(e) shows that our deconvolution procedure can figure out those objects with no crosstalk. In terms of the imaging resolution, deconvolution result in  Figs. \ref{fig7}(a) and  \ref{fig7}(b) shows that the spatial resolution of our system reaches 0.15mm at 500nm, while the diffraction limit of our system at the OP is 0.134mm at 500nm, calculated by the diameter of the fiber bundle.

\begin{figure*}[!t]
\centering
{\includegraphics{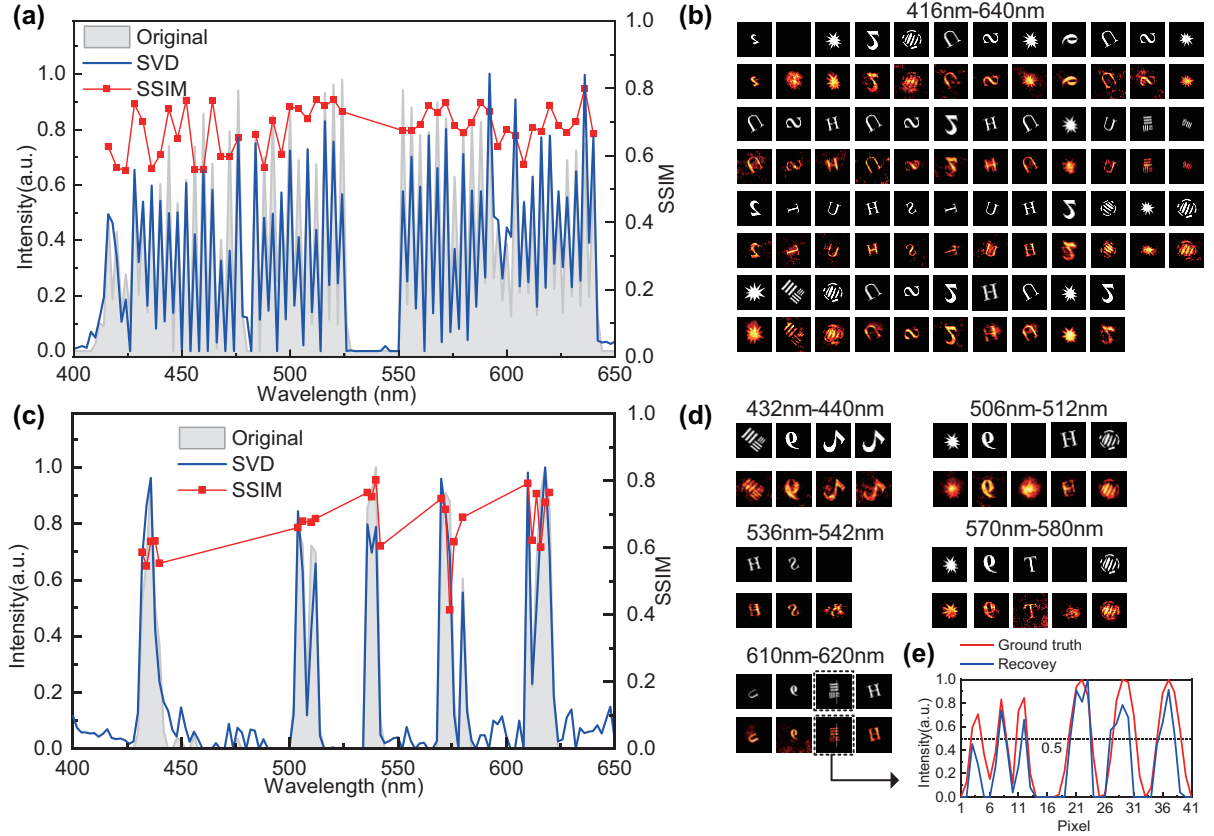}}
\caption{Hyperspectral imaging for complex objects. (a) Original and recovered spectrum at ${{90}^{\circ}}$polarization state. (b) The ground truth and denoised images corresponding to channels in (a). (c) Original and recovered spectrum at the ${{0}^{\circ}}$ polarization state. (d) The ground truth and denoised images  corresponding to channels in (d). Each SSIM index of the object channel is also plotted as the red square in (a),(c). (e) The intensity profile of the resolution target along the white dashed line at 618nm at ${{0}^{\circ}}$ polarization state. It shows our imaging resolution reaches 0.16mm at 618nm, close to the diffraction limit.}
\label{fig8}
\end{figure*}

We ensure the continuity of spectral information in $T_c$ by setting suitable calibration linewidth and sampling interval . As a result, the continuous spectrum from 504nm to 512nm is successfully recovered, and the continuity of the spectrum does not affect our system performance directly. However, the spectrum's sparsity does affect the complexity of the information and the SNR of the speckle, which further affects our system performance.

\subsection{Complex Objects with Various Dimensions}

Here we perform our hyperspectral imaging method in a low SNR situation. The target object consists of figures, letters with their diameters varying from 1.3mm to 3mm. In the 90$^\circ$ polarization state, the object's channels have 4nm spectral spacing and 2nm linewidth to form the jagged ground truth spectrum. In its orthogonal polarization state, the objects have a continuous spectrum at different central wavelengths. We also set some background channels to test the spectral-spatial resolution and crosstalk of the system.

Figure \ref{fig8} plots the recovery result. In the 90$^\circ$ polarization state, our method positions each object's channel with high wavelength accuracy to recover the jagged original spectrum. However, the spectral intensity is inaccurate in some channels. The recovery spectrum matches the ground truth in its orthogonal polarization, but background noise also appears. To further explain the recovery quality, we calculate the correlation value between the ground truth and the recovery spectrum. The result turns out to be 0.9403, meaning that the recovered spectrum matches well with the original object on the whole.

 In this situation, noise as well as calibration mismatch mainly affects the PSNR of the spectrum, while the wavelength positioning remains reliable.  Theoretically, the speckle is composed of $PSF(\lambda_i)$ and additive noise. In the recovery procedure, the additive noise tends to be randomly allocated to all spectral channels, causing background noise and intensity inaccuracy, and $PSF(\lambda_i)$ components would be allocated to its original spectral channel at $\lambda_i$. As a result, low SNR results in degraded recovery quality.

The imaging result is shown in Fig. \ref{fig8}. The deconvolution procedure can explain the degradation in imaging quality: as we deconvolute $I$ with $PSF(\lambda_i)$, channels that do not belong to $\lambda_i$ become background noise. To further explain the recovery quality, we calculate the SSIM between the recovered object and the original. In the spectral channel where the intensity at the OP is relatively low, the SSIM also drops to a lower value. However, the crosstalk between neighborhood channels is still low enough to achieve 2nm resolution, which can be seen in Fig. \ref{fig8}(d). We also test the imaging resolution by calculating the profile of the resolution target at 618nm. The result plotted in Fig. \ref{fig8}(e) shows that our system achieves the resolution of 0.16mm at 618nm, which is very close to the diffraction limit (0.166mm at 618nm). However, we also point out that the Wiener deconvolution algorithm requires a higher Noise-to-Signal Ratio (NSR) index to recover complex objects, which will result in the loss of imaging resolution.

\subsection{Color Imaging in Heavily Mismatched Situation}
\label{sec3e}
\begin{figure}[tb!]
\centering
{\includegraphics[width=0.7\linewidth]{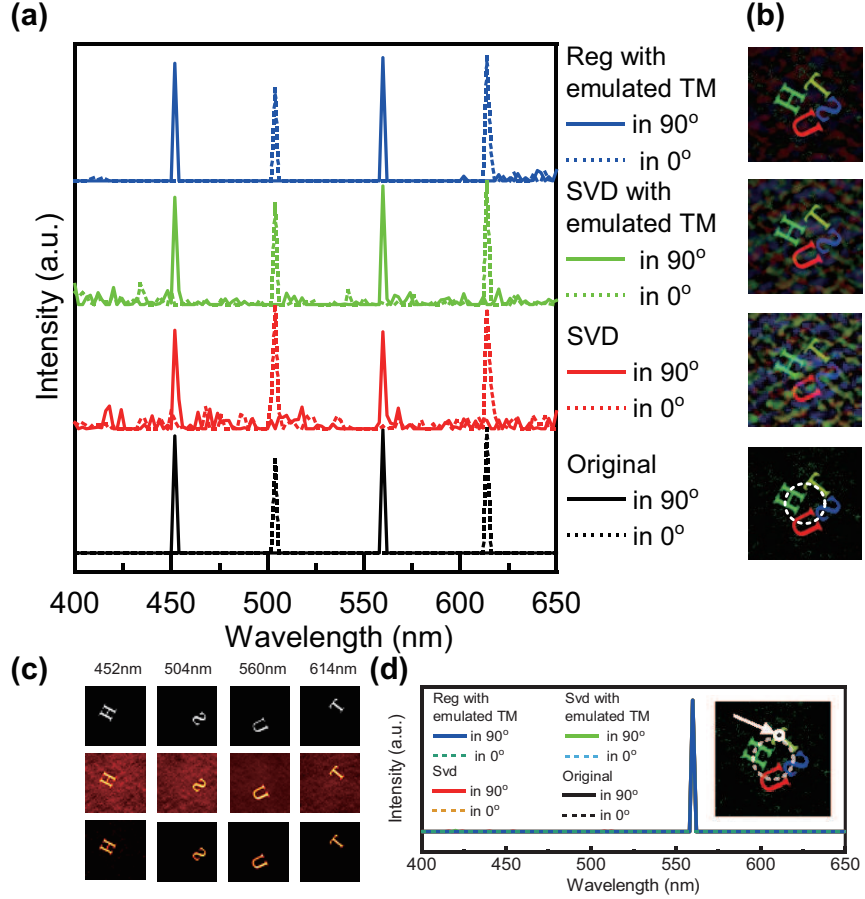}}
\caption{
(a) The ground truth (bottom) and the recovered spectrum from different algorithms. The 'Reg' on the top denotes the nonlinear regulation optimization.  (b) The ground truth (bottom) and the recovered color object from different algorithms. The recovered spectrum's contrast will affect the imaging quality of the recovered object. The Nonlinear optimized method, combined with $T_e$, matches the ground truth best. (c) The images of the true object (top), the recovered object with noise (middle), and the denoised object (bottom) in each spectral channel. (d) The spectrum for a pixel at the letter 'T'. The white dot and arrow label the pixel. The white dashed line labels the size of the calibration aperture in this experiment.}
\label{fig9}
\end{figure}

To overcome the mismatch between the calibration aperture and the hidden object, we reconstruct the TM with the emulation method to conduct a color imaging experiment. Here we set a series of target objects, consisting of the letter 'H',' U', 'S',' T' with 1.3mm in diameter. To enlarge the mismatch between the objects and the calibration aperture, we shift objects away from the aperture's center. The calibration aperture is plotted as a white dashed line in Fig. \ref{fig9}(d). For each object, only approximately 50\% of its structure is inside the aperture area, and the SSIM between the objects and the calibration aperture is 0.23 on average.

While the object obviously mismatches the aperture, the SVD recovery result shows that the blurred speckle's overlaid structure remains highly wavelength dependent. However, we also notice that high mismatch also causes reconstruction error. For example, the relative intensity is inaccurate compared with the ground truth, and noise occurs in background channels. We try a nonlinear optimization to correct the error, but the result is not satisfying because the optimized target $\left\| I-T_c\cdot S' \right\|_{2}^{2}+{{\gamma }_{1}}{{\left\| S' \right\|}_{1}}$ mismatches to the real $T$. In this situation, the reconstruction error mainly comes from the model error in our approximation $T_c\approx{T}$ rather than the speckle's SNR. As a result, we try another way to reconstruct the TM by the method discussed in Section  \ref{secII}. We firstly calculate $T_e$ with $O_d$ by Eq. \ref{eq:eq13}, then use the same procedure to recover the spectrum with $T_e$. The recovery in Fig. \ref{fig9}(a) shows better performance in intensity accuracy. Furthermore, we conduct the nonlinear regulation optimization with a target of $\left\| I-T_e\cdot S' \right\|_{2}^{2}+{{\gamma }_{1}}{{\left\| S' \right\|}_{1}}$. The result is quite satisfying to meet with the ground truth.

We also conduct the imaging method in each channel. Each object as well as its recovery is plotted in Fig. \ref{fig9}(c). We plot each channel by the spectral color of its wavelength in CIE 1931 RGB space \cite{Smith_1931} and then compose them into a color image with their spectral intensities as coefficients. Figure \ref{fig9}(b) shows the color image from each reconstruction algorithm as well as the ground truth. As we can see, the color image from $T_e$ shows better contrast than $T_c$'s result, and the nonlinear regulation algorithm further improves the imaging quality. Then we calculate the spectral information of a single-pixel at the letter 'T', which is plotted as a white dot in Fig. \ref{fig9}(d). For different recovery algorithms, the spectral intensity of the target pixel always matches the ground truth well. This is because noise components tend to be distributed equally into the whole OP, but in the object's channel, the intensity component gathers together to form the object with high contrast. As a result, system noise mainly affects background pixels. Based on this feature, further algorithms are practicable to perform denoising for our hyperspectral imaging method.

\section{DISCUSSION}

%Our method has shown its feasibility of recovering the complex hyperspectral information inside the speckle without any special optical design or element.
In our proposed method, we successfully apply the spectral intensity TM approach to retrieve the spatial and hyperspectral information from scattered incoherent light with different polarization states, which, however, has once been regarded as fiendishly complex or even impractical in former studies. %It further reveals the potential to explore the abundant, scrambling information in incoherent light spots, which, however, has once been regarded as a form of information loss. As a result, we prove that the principle of light scattering itself, rather than the technical extension of other optical elements, can be  %taken advantage of to be a new access to the hyperspectral information in natural situations.
Moreover, compared to traditional methods by coupling coherent light into a single-mode polarization-maintaining fiber, the utilization of spatially incoherent light in our proposed method can further enhance the system's adaptability to various input of spatial distribution. %compared with coherent light coupled into a single-mode polarization-maintaining fiber, spatially incoherent light enhance the system's stability for the retrieval of input spatial information.
Our simulation result shows that the spectral retrieval remains high quality for various objects, even for a single-pixel object, as long as the object is inside the area of the calibration aperture.

 %While the light scattering in natural scenes used to be regarded as a form of information loss, we offer the access to take advantage of the scattering itself hyperspectral information encoded in the blurred light spot.
 %上面这句话那个更好？

%In our proposed method, the spectral TM theory is successfully extended to the incoherently illuminated objects.
%In our proposed method, we successfully apply the spectral intensity TM to retrieve objects under incoherent illumination. However, using the highly incoherent light with different polarization states as the light source was regarded as fiendishly complex or even impractical in former studies.

We put forward a method to deal with the calibration mismatch for objects out of the calibration area. The main restriction is that the emulated convolution procedure to calculate $T_e$ requires high SNR of PSFs in each channel. We have managed to achieve a better result with PSFs by deconvoluting the aperture's speckle with the aperture's binary image because the digital deconvolution and convolution procedure effectively reduces additive noise. The calibration procedure can also be simplified via utilizing PSFs to emulate the aperture's speckle or utilizing the aperture's speckle to calculate PSFs by deconvolution.

%Another attempt is to calibrate the spectral TM with the linewidth broader than the medium's spectral correlation width—this idea is originated from the low spectral energy density in natural scenes. So we try to merge the narrow-linewidth spectral channels (assumed to be 0.05nm) to create the broad-linewidth spectral channels with higher SNR per channel. This attempt successfully performs the spectrum retrieval for various light sources from the narrow-spectrum lasers to the broad-spectrum light sources. More importantly, this method shows the ability to control the system's spectral resolution by adjusting the channel linewidth and spacing. This improves the flexibility to vary the system's spectral resolution without replacing the medium itself. Using the broad-linewidth spectral channels at the spectral resolution cost also improves the system's robustness over temperature variations and other environmental disturbances. As a result, by using multiple TMs calibrated by different linewidth, our system can obtain the high-resolution spectrum and the high-SNR spectrum from a single measurement.

As is discussed in Section \ref{sec:Intro}, the fiber bundle in our experiment has good temporal stability, large memory effect feature, and high spectral sensitivity. Use of the fiber bundle enables us to easily calculate the memory effect range, spectral bandwidth, and other system properties through the bundle's parameters. As a general method, our technique can be further adopted in other disordered media with light scattering features similar to the fiber bundle.

The memory effect range is measured to be 35mrad FWHM. Exceeding the memory effect range to perform calibration without any speculation of its dimension is also possible by applying Eq. \ref{eq:eq13} and $O_n$ to reconstruct the TM in large FOV. By far, hyperspectral imaging methods for large FOV are currently under our investigation.

While the incoherent situation frees the optical system from interference measurement, it also degrades the speckle to a much lower SNR. In our experiments, SNR is the main factor to determine the spectral recovery performance. Although our fiber bundle provides lots of spectral channels (assumed to be more than 50,000), the speckle SNR limits the target object's complexity.

The speckle SNR also affects the imaging procedure. While the Wiener deconvolution algorithm recovers the spatial information at $\lambda_i$ with $PSF(\lambda_i)$, other spectral channels become background noise and cause low contrast of the recovered image. As a result, the denoising procedure in our method is necessary to improve the image's contrast. Many methods have been put forward to solve the denoising problem in hyperspectral imaging \cite{dabov2007image,fu2015adaptive}. However, the noise mechanism in the deconvolution-based hyperspectral imaging method is different from conventional methods, so further algorithms to solve the deconvolution problem in heavy spectral complexity and low SNR situation as well as the denoising methods, are remaining to be explored. In our study, adjusting the NSR index of Wiener deconvolution algorithm from $10^6$ (Section \ref{sec3a}) to $2*10^9$ (Section \ref{sec3e}) can satisfy our imaging demands at the cost of imaging resolution. Besides, deep-learning \cite{xie2017hyperspectral} and compressed sensing \cite{French:18} techniques can also be adopted with our method to improve recovery quality.

%Some attempts, including the calibration mismatch in $T_c$ and broad-linewidth calibration, are unnecessary when each channel's SNR is relatively high, as they lead to model errors in the recovery.
To deal with model errors, we apply several methods to improve the robustness of our system. For example, the speckle filtering algorithm is essential to retrieve the target spectrum, especially for complex spectrum. We also find that, although the detected speckle is only a subcomponent of the superposed speckle in $T_c$ or $T_e$, the matrix inversion and the nonlinear regulation procedure handle those model errors well. In this situation, for a signal channel, its wavelength position remains highly accurate, but part of its spectral intensity is turned into background noise to compensate for speckle's lost subcomponents. Due to the large number of spectral channels in our system, the reconstructed spectrum is still reliable on the whole, even for a point object. In our experiments, model errors are acceptable for most applications that do not require accurate spectral intensity retrieval, and we can improve the system SNR to avoid those model errors.

The system SNR can be improved by using a broadband tunable laser with a rotating ground glass as the calibration light source. Another option is utilizing a dichroic mirror to separate the broadband speckle into multiple detection channels, each receiving a subset of the input spectrum. All detection channels can still be instrumented with the same fiber bundle as the dispersion element to sufficiently reduce cost and volume.

%系统SNR与Model error相互制约

Theoretically speaking, the imaging crosstalk between spectral channels can be solved by conducting the nonlinear optimization to minimize $\left\| I-\sum{O({{\lambda }_{i}})*PSF({{\lambda }_{i}})} \right\|$ \cite{antipa2018diffusercam}, but the optimization requires high memory storage of the computer and long run time for our large bandwidth system. In our experiments, the spectral recovery takes 0.002s. Our deconvolution-based hyperspectral imaging in 252 channels takes only 44.8s, while the nonlinear algorithm is out of memory on a GeForce GTX 1660 GPU.

\section{CONCLUSION}
Our work presents a new polarized hyperspectral imaging method in incoherent situation with a fiber-bundle-only design. While the object’s spatial distribution varies, the seemly random speckle is actually the overlay of its wavelength feature, $PSF(\lambda)$, making the speckle hold strong features of wavelength-dependence. We extend the spectral intensity TM from coherent light in a single-mode polarization-maintaining fiber to an extended object under incoherent illumination to adapt to common natural scenes. We then establish the convolution imaging approach to compose the TM. This idea enables us to separate the spectral information from the hyperspectral cube by matrix inversion. Based on the knowledge of the similar overlapping structure of $PSF(\lambda)$, we put forward the solution to approximate the unknown TM with an experimentally calibrated or digitally emulated result. We demonstrate our method by building the hyperspectral imaging system with a fiber bundle in the range of 400nm-650nm for two polarization states with 2nm spectral channel spacing. Then we perform the reconstructions in different spatio-spectral scenes to discuss the quality of imaging and spectral recovery. Our system solves the light scattering problem for extended objects under incoherent illumination. It is cheap, concise, and does not rely on any extra imaging, polarization, or dispersion element except the fiber bundle itself. Our system can be applied to other disordered media, can be further extended to large FOV situations, and can also be combined with compressed sensing or deep learning to improve system performance.

\medskip

\noindent\textbf{Funding.} National Natural Science Foundation of China (61805088); Science, Technology, and Innovation Commission of Shenzhen Municipality (JCYJ20190809100811375); Key Research and Development Program of Hubei Province (2020BAB121); Fundamental Research Funds for the Central Universities (2019kfyXKJC040); Innovation Fund of WNLO.

\medskip

\noindent\textbf{Disclosures.} The authors declare no conflicts of interest.

\medskip

\bibliographystyle{unsrt}
\bibliography{refs}

% Full bibliography added automatically for Optics Letters submissions; the following line will simply be ignored if submitting to other journals.
% Note that this extra page will not count against page length



%\begin{thebibliography}{1}

%\bibitem{kour2014real}
%George Kour and Raid Saabne.
%\newblock Real-time segmentation of on-line handwritten arabic %script.
%\newblock In {\em Frontiers in Handwriting Recognition (ICFHR), 2014 14th
%  International Conference on}, pages 417--422. IEEE, 2014.

%\bibitem{kour2014fast}
%George Kour and Raid Saabne.
%\newblock Fast classification of handwritten on-line arabic characters.
%\newblock In {\em Soft Computing and Pattern Recognition (SoCPaR), 2014 6th
%  International Conference of}, pages 312--318. IEEE, 2014.

%\bibitem{hadash2018estimate}
%Guy Hadash, Einat Kermany, Boaz Carmeli, Ofer Lavi, George Kour, and Alon
%  Jacovi.
%\newblock Estimate and replace: A novel approach to integrating deep neural
%  networks with existing applications.
%\newblock {\em arXiv preprint arXiv:1804.09028}, 2018.

%\end{thebibliography}

\end{document}